\documentstyle[12pt]{article}
\input{psfig}
\begin{document}

\begin{flushright}
        \small
        hep-ph/9803460\\
        JUNE 1998
\end{flushright}

\begin{center}
\vspace{1cm}
{\large \bf The Promising Process to Distinguish Supersymmetric Models with Large tan$\beta$ from the 
Standard Model: $B\rightarrow X_s{\mu}^{+}{\mu}^{-}$}

\vspace{0.5cm}

\renewcommand{\thefootnote}{\fnsymbol{footnote}}
{ \bf Chao-Shang Huang$^a$\footnote{\it csh@itp.ac.cn}, Wei Liao$^b$\footnote{\it liaow@itp.ac.cn}, and Qi-Shu Yan$^c$\footnote{\it yanqs@itp.ac.cn}\\
Institute of Theoretical Physics, Academia Sinica.\\
P. O. Box 2735, Beijing 100080, P. R. China}

\vspace{.8cm}
{\large \bf abstract}
\end{center}
\vspace{0.5cm}
It is shown that in supersymmetric models (SUSYMs) the large supersymmetric contributions to $B \rightarrow X_s{\mu}^{+}{\mu}^{-}$
come from the Feynman diagrams which consist of exchanging neutral Higgs bosons
(NHBs) and the chargino-stop
loop and are proportional to $m_b m_{\mu}$tan$^3\beta/m_{h}^2$ when tan$\beta$ is large
and the mass of the lightest neutral Higgs boson m$_h$ is not too large (say, less than
150 Gev). Numerical results show that the branching ratios of $B \rightarrow X_s{\mu}^{+}{\mu}^{-}$ 
can be enhanced by more than 100\% compared to the standard model (SM) and the backward-forward asymmetry
of lepton is significantly different from that in SM when tan$\beta \!\geq \!30$.

\vskip 0.5cm
 
\indent It is widely believed that supersymmetry (SUSY) is one of the most promising candidates for physics
beyond SM since it offers a scheme to embed the SM in a more fundamental
theory in which many theoretical problems such as gauge hierarchy, origin of
mass and Yukawa couplings can be answered. One direct way to search for SUSY
is to discover SUSY particles at colliders. But, unfortunately, so far
no SUSY particles have been found. Another way is to search for its effect through
indirect methods. In most of SUSYMs R parity is conserved so that SUSY
contributions to an observable appear at the loop level. Therefore, it has been
realized for a long time that rare processes can be used as a good probe
for searches of SUSY, since in these processes the contributions of SUSY
and SM arise at the same order in perturbation theory. The $B\rightarrow X_s l^{+}l^{-}$ (l=e, $\mu$, $\tau$)
process, one of rare processes, in SUSYMs has been extensively studied [1--5]. The effects of large
tan$\beta$ have been noticed in recent papers \cite{bll1, bll2}. There is a stop-chargino
loop diagram which gives a large contributions to $C_7$ when tan$\beta$ is large \cite{bsg4}.
This leads to that in the minimal supergravity model (mSUGRA) there are regions in
the parameter space where the branching ratio of $b\rightarrow s l^{+}l^{-}$ (l=e, $\mu$)
is enhanced by about 50\% compared to the SM \cite{bll2}. However, the contributions
from exchanging NHBs are ignored in these previous analyses.
Recently, the contributions of NHBs in SUSYMs have been taken into account.
Because the contributions to
$b\rightarrow s \tau^{+}\tau^{-}$ coming from the chargino-stop loop diagram are proportional to $m_b m_{\tau}$ tan$^3 \beta/m_h^2$
$(h=h^0, A^0)$ when tan$\beta$ is large, the branching ratio of $b\rightarrow s \tau^{+}\tau^{-}$
can be enhanced by about 200\% compared to the SM \cite{bll3}.

From experimental points of view, the observation of $B\rightarrow X_s l^{+}l^{-}$ (l=e, $\mu$)
is easier accessible than that of $B\rightarrow X_s \tau^{+}\tau^{-}$.
The inclusive decay $B\rightarrow X_s \gamma$ has been observed by CLEO.
Meantime, experiments at $e^{+} e^{-}$ and hadron colliders are closing in on the 
observation of $B\rightarrow K^{*} l^{+} l^{-}$ (l=e, $\mu$) \cite{exp1}. The B factories
presently under construction will collect some $10^7$--$10^8$ B mesons
per year which can be used to obtain good precision on low branching
fraction modes. Therefore, it is meaningful to
pay attention to the process $B\rightarrow X_s l^{+}l^{-}$ for (l=e, $\mu$).
As pointed above, the contributions of NHBs
are proportional to the mass of a lepton and tan$^3 \beta$.
For $B\rightarrow X_s e^{+} e^{-}$,
the contributions can be safely neglected due to the smallness of $m_e$, no matter
how large tan$\beta$ is (of course, in the theoretically allowed range, say, in SUSYGUT, tan$\beta\!\leq\! 50$).
However, for $b\rightarrow s \mu^{+}\mu^{-}$, $m_{\mu}$ tan$\beta$ can be as large
as $m_{\tau}$ as long as tan$\beta \! \geq \! 17$. Thus one can expect that for $b\rightarrow s \mu^{+}\mu^{-}$,
in addition to the enhancement coming from the possible change of the sign of $C_7$ (
the value of $C_7$ is fixed by the measurement of $b\rightarrow s \gamma$ with a
branching ratio of $(2.32\pm0.57\pm0.35)\times 10^{-4}$ and $95$\% C.L. bounds
of $1 \times 10^{-4} <Br(B \rightarrow X_s \gamma)<4.2\times 10^{-4}$ \cite{exp2}), a even more
significant enhancement coming from exchanging NHBs arises in SUSYMs with large
tan$\beta$. In the letter we calculate the invariant mass distribution and
backward-forward asymmetry of dilepton angular distribution for $B\rightarrow X_s \mu^{+}\mu^{-}$
in SUSYMs. Our results show that the branching ratio of $B\rightarrow X_s \mu^{+}\mu^{-}$
is enhanced by at least 100\% compared to the SM and the back-forward asymmetry
is more sensitive to tan$\beta$ than the invariant mass distribution when tan$\beta$
is larger than 30 and masses of Higgs bosons, stops and charginos are in the
reasonable range (i.e., all constraints from phenomenology are satisfied). Note that
the invariant mass distribution of $B\rightarrow X_s \mu^{+}\mu^{-}$ in a two Higgs
doublet model with large tan$\beta$ is not enhenced compared to the SM. 
Therefore, the rare process $B\rightarrow X_s \mu^{+}\mu^{-}$ provides a good opportunity
to distinguish SUSYMs with large tan$\beta$ from the SM and
it is possible that the first distinct signals of SUSY could come from
deviations from the SM in the inclusive decay $B\rightarrow X_s \mu^{+}\mu^{-}$.

Inclusive decay rates of $B\rightarrow X_s {\mu}^{+}{\mu}^{-}$ can be 
calculated in the $1/m_Q$ expansion and it has been shown that the leading 
order term turns to be the decay of a free b quark and corrections stem from 
the $1/m_Q^{2}$ order and are small about a few percent \cite{hqet}. Therefore in what
follows we limit our analyses to the leading order term, i.e.\ the decay 
$b\rightarrow s{\mu}^{+}{\mu}^{-}$. 

There are several classes of new contributions in SUSYMs and the dominant ones are 
arising from chargino(${\tilde{\chi}}$)-uptype squark loop and charged Higgs boson ($H^{\pm}$)-uptype
quark loop. Because of small generation mixing of squarks coming from phenomenological 
constraints on $K^{0}-{\bar {K^0}}$ and $D^{0}-{\bar {D^0}}$,
the contributions from gluino-downtype squark loop and neutralino-downtype squark
loop are much smaller than the dominant ones and are neglected in the following.

The effective Hamiltonian relevant to 
the $b\rightarrow s\mu^{+}\mu^{-}$ process is 
\begin{eqnarray}
H_{eff} & = & \frac{4G_{F}}{\sqrt{2}} V_{tb} V_{ts}^{*} (\sum_{i=1}^{10}
    C_{i}(\mu) O_{i}(\mu) + \sum_{i=1}^{10} C_{Q_i}(\mu) Q_i(\mu))
\end{eqnarray}
where $O_i$ (i=1, 2, ..., 10) are given in Ref.\ \cite{bll4}, and $Q_i$'s come from exchanging 
neutral Higgs bosons and have been given in Ref.\ \cite{bll5}. The coefficients $C_i(m_w)$ in 
SUSYMs have been calculated \cite{bsg1, bsg3}. We calculate
the coefficients $C_{Qi}(m_W)$ in SUSYMs and the results are:
\begin{eqnarray}
C_{Q1}(m_W) &=& \frac{m_b m_{\mu}}{4 m_{h^0}^2 \sin^2\theta_W} {\rm tg}^2 \beta \{
 (\sin^2\alpha + h \cos^2\alpha) [ \frac{1}{x_{Wt}}(f_1(x_{Ht})-f_1(x_{Wt}))\nonumber\\
   &&+\sqrt{2} \sum_{i=1}^{2} \frac{m_{\chi_{i}}}{m_W} \frac{U_{i2}}{\cos \beta} (- V_{i1} f_1(x_{\chi_{i}{\tilde q}})
    +\sum_{k=1}^{2}\Lambda(i,k)T_{k1}f_1(x_{\chi_{i} {\tilde t_k}}))\nonumber\\
   &&+(1+\frac{m_{H_{\pm}}^2}{m_W^2}) f_2(x_{Ht},x_{Wt})]   
    -\frac{m_{h_0}^2}{m_W^2} f_2(x_{Ht},x_{Wt})\nonumber\\
   &&+2 \sum_{ii'=1}^2 (B_{1}(i,i') \Gamma_{1}(i,i')+A_{1}(i,i') \Gamma_{2}(i,i'))\}\nonumber\\
C_{Q2}(m_W) &=&- \frac{m_b m_{\mu}}{4 m_{A^0}^2 \sin^2\theta_W}{\rm tg}^2 \beta\{\frac{1}{x_{Wt}}(f_1(x_{Ht})-f_1(x_{Wt}))
    +2 f_2(x_{Ht},x_{Wt})\nonumber\\
    &&+ \sqrt{2} \sum_{i=1}^{2} \frac{m_{\chi_{i}}}{m_W} \frac{U_{i2}}{\cos \beta} (- V_{i1} f_1(x_{\chi_{i} {\tilde q}})
    +\sum_{k=1}^{2}\Lambda(i,k)T_{k1}f_1(x_{\chi_{i}{\tilde t_k}})) \nonumber\\
    &&+ 2 \sum_{ii'=1}^2 (-U_{i'2}V_{i1} \Gamma_{1}(i,i')+U^{*}_{i2}V^{*}_{i'1} \Gamma_{2}(i,i'))\}
\end{eqnarray}
where the definitions of the functions $f_i$, $\Gamma_i$ (i=1, 2), $A_1$, $B_1$,
$\Lambda$ and the meaning of the matrices $U$, $V$, $T$ have been given in \cite{bll3}
and we have omitted less important terms because they are numerically negligible compared
to those given in eq. (2) when tan$\beta \geq 20$.

From eq. (2), we see that, for large tan$\beta(\geq\! 20)$, the coefficients
$C_{Q_i}$ are proportional to ${m_b m_{\mu}}$/$m_{h}$tan$^3 \beta$ $(h=h^0, A^0)$. One 
factor of tan$\beta$ comes from the chargino-up-type squark loop and tan$^2 \beta$ from 
exchanging the neutral Higgs bosons (note that in the large tan$\beta$ approximation
cos$^{-1}\beta\simeq$tan$\beta$). Therefore, $C_{Q_i}$ can compete with $C_i$ 
and even overwhelm $C_i$ as long as tan$\beta$ is large enough. We remark that the 
chirality structure of the $Q_i$ (i=1, 2) operators allows a large tan$\beta$ enhancement
for the $C_{Q_i}$ (i=1, 2) coefficients, as happened for the magnetic moment operator O$_7$,
and there is no such a large tan$\beta$ enhancement for the $C_i$ (i=8, 9) coefficients
due to the different chirality structure of the $O_i$ (i=8, 9) operators.
Incorporating the QCD corrections to the coefficients $C_i$ and $C_{Q_i}$ in
the standard way, we calculate these coefficients at $\mu$=$m_b$.

The differential branching ratio and the forward-backward 
asymmetry of the dimuon angular distribution for $B\rightarrow X_{s} \mu^{+}\mu^{-}$
can be obtained from \cite{bll3} with substituting $m_{\mu}$ for $m_{\tau}$.
The numerical results of the invariant mass distribution and backward-forward asymmetry
are shown in Fig.\ 1 (a) for the mSUGRA model and Fig.\ 1 (b) for the MSSM with a
typical choice of masses of sparticles and Higgs bosons respectively. The
mSUGRA parameters ($m_0$, $m_{1/2}$, A)=(190, 190, 380) Gev, Higgs mass mixing parameter
$\mu \! < \! 0$ and tan$\beta$=30 have 
been chosen in Fig.\ 1 (a). In the computations of sparticle mass spectra and mixings
we neglect the Yukawa couplings of the first two generations. The chosen values of masses of relevant sparticles and 
Higgs bosons in MSSM are given in the Figure Captions. The constraint from the LEP and 
$b\rightarrow s\gamma$
has been imposed in our numerical calculations. One can see from the 
Fig.\ 1 that a large enhancement of the differential branching ratio 
$d{\Gamma}/{ds}$ shows up and the enhancement can reach $100\%$ compared to SM
when tan$\beta$=30. The backward-forward is significantly different from that in
SM. The predictions without including the contributions of exchanging NHBs are
also shown in Fig.\ 1 in order to compare. It is evident from the figure
that the contributions of exchanging NHBs to the differential branching ratio are the same
order of magnitude as supersymmetric contributions without including exchanging NHBs
in the low s region ($s{\leq}0.4$) and larger than those in the high s region ($s\!>\!0.4$). There are
regions in the parameter space where the contributions of NHBs alone make a
large enhancement
of the differential branching ratio. For example, for a set of
values of parameters
($\theta_{\tilde {t}}$=$-20$$^{\circ}$, $m_{\chi_2}$=220 Gev,
$m_{\chi_1}$=100 Gev, $m_{\tilde {q}}$=430 Gev, $m_{{\tilde t}_{1}}$=250 Gev, $m_{{\tilde t}_{2}}$=500 Gev,
$m_{A^0}$=80 Gev, $m_{\tilde nu}$=160 Gev, and tan$\beta$=30) the enhancement of
$d \Gamma$/$d s$ coming from NHBs is about $80\%$ compared to SM.

We would like to make some remarks: 

\indent (i) The large enhancement of the invariant mass distribution of $B\rightarrow 
X_s{\mu}^{+}{\mu}^{-}$ compared to the SM is of a common feature of SUSYMs
with large tan$\beta$ in some region of the parameter space. The Fig.\ 2
shows the results for tan$\beta$=30. For larger tan$\beta$, for example,
tan$\beta$=45, the enhancement can reach $200\%$. The enhancement exists as long 
as the mass splitting of stops is large enough (say, ${\geq}$100 Gev).
The condition is necessary because if all the squark masses are degenerate ($m_
{\tilde{t}_1}=m_{\tilde{t}_2}=\tilde{m}$), the large contributions arising from
the chargino-squark loop exactly cancel due to the GIM mechanism \cite{bsg4}. In order to illustrate this point we show the $C_{Q_i}$ as a function of the 
stop mixing angle ${\theta}_{\tilde t}$ under the above condition
in Fig.\ 2. As can be seen from the Figure, $C_{Q_i}$ is large enough to make an 
large enhancement of ${d{\Gamma}}/{ds}$ in a wide range of $\theta_{\tilde t}$ (about
from -$\frac{2 \pi}{5}$ to -$\frac{\pi}{8}$).
 
(ii) The QCD corrections to coefficients $C_i$ and $C_{Q_i}$ are incorporated in 
the leading logarithmic approximation in our numerical computations. 
The one-loop mixing of $Q_i$ with $O_7$ has been analyzed \cite{bll5} and 
leads to about $5\%$ correction to
$C_7(m_b)$ when tan$\beta$=30. A next-to-leading order(NLO) 
analysis without
including $Q_i$ for $B\rightarrow X_s l^{+}l^{-}$ has been performed, where it 
is stressed that a scheme independent result can only be obtained by including 
the LO and NLO corrections to $C_8^{eff}$ while retaining only the LO corrections
in the remaining Wilson coefficient \cite{bll6}. Because we did not include the NLO corrections
the theoretical uncertainty due to the renormalization scale ${\mu}$ dependence
is about ${\pm}20\%$ as ${\mu}$ is varied in the range $1/2$ $m_b{\leq}{\mu}{\leq}2m_b$.
We expected that a full NLO analysis including $Q_i$ for $B{\rightarrow}X_sl^{+}l^
{-}$ will appear in the near future. As pointed in the Ref.\ cite{bsg5}, there is a SUSY high 
scale uncertainty and it is possible that the scale ${\mu}$ dependence is large
enough to effectively encompass the uncertainty.

(iii) The following values of parameters have been used in the numerical calculations:
$m_t=175 Gev$, $m_c/m_b=0.3$, ${\eta}{\equiv} {{\alpha}_s(m_b)}/{{\alpha}_s(m_w)}=0.548$.
We have estimated the uncertainties from the parameters and  results are that the 
$m_t$ dependence is weak and the uncertainties are about ten percent. The error from neglecting the strange quark
mass $m_s$ is of order ${m_s^2}/{m_b^2}$ and consequently is very small.

In summary, we have investigated the differential branching ratio and
backward-forward asymmetry of lepton for $B{\rightarrow}X_s{\mu}^{+}{\mu}^{-}$ in SUSYMs
with large tan$\beta$. There is a $100\%$ enhancement of the differential branching
ratio compared to SM if tan$\beta{\geq}30$ and the masses of Higgs bosons,
squarks and charginos are in the reasonable range. Because there is almost no 
enhancement till tan$\beta$=50 in a two Higgs doublet, one can make the conclusion
that the first distinct signals of SUSY could come from the observation of
$B \rightarrow X_s{\mu}^{+}{\mu}^{-}$.

This research was supported in part
by the National Natural Science Foundation of China and partly supported by
Center of Chinese Advanced Science and Technology(CCAST).

\vskip 1cm
\begin{center}
{\Large \bf Figure Captions}
\end{center}
\vskip 0.5cm
\noindent Fig.\ 1 $d\Gamma/ds$ and A(s) for the case $\mu\!<\!0$ and tan$\beta$=30,
a) $m_{1/2}=m_0=190$ Gev, $A=380$ GeV in the mSUGRA and b) $\theta_{\tilde {t}}$=$-40$$^{\circ}$,
$m_{\chi_2}$=200 Gev, $m_{\chi_1}$=90 Gev,
$m_{\tilde {q}}$=350 Gev, $m_{{\tilde t}_{1}}$=220 Gev, $m_{{\tilde t}_{2}}$=450 Gev,
$m_{A^0}$=80GeV, $m_{\tilde {\nu}}$=160 Gev in the MSSM. 
The solid , dashed and dotted lines represent the predictions of the
SUSYMs, the SUSYMs without including contributions of NHBs and
SM respectively.

\vskip 0.5cm
\noindent Fig.\ 2 $C_{Q1}$ and $C_{Q2}$ varying with the mass splitting of stop, 
the mass splitting of chargino and the 
stop mixing angle ${\theta}_{\tilde t}$ for $\mu\!<\!0$, tan$\beta$=30,
$m_{{\tilde t}_{1}}$=150 Gev, $m_{\chi_1}$=90 Gev and $m_{A^0}=80$ Gev; the characters following the line style indicate
the mass splittings: the first character means
the mass splitting of stop (h represents 300 Gev and l 100 Gev), 
and the second character means that of chargino (h represents 410 Gev,
m 210 Gev, and l 110 Gev).

\newpage
\begin{figure}[t]
\vspace{0cm}
\centerline{
\psfig{figure=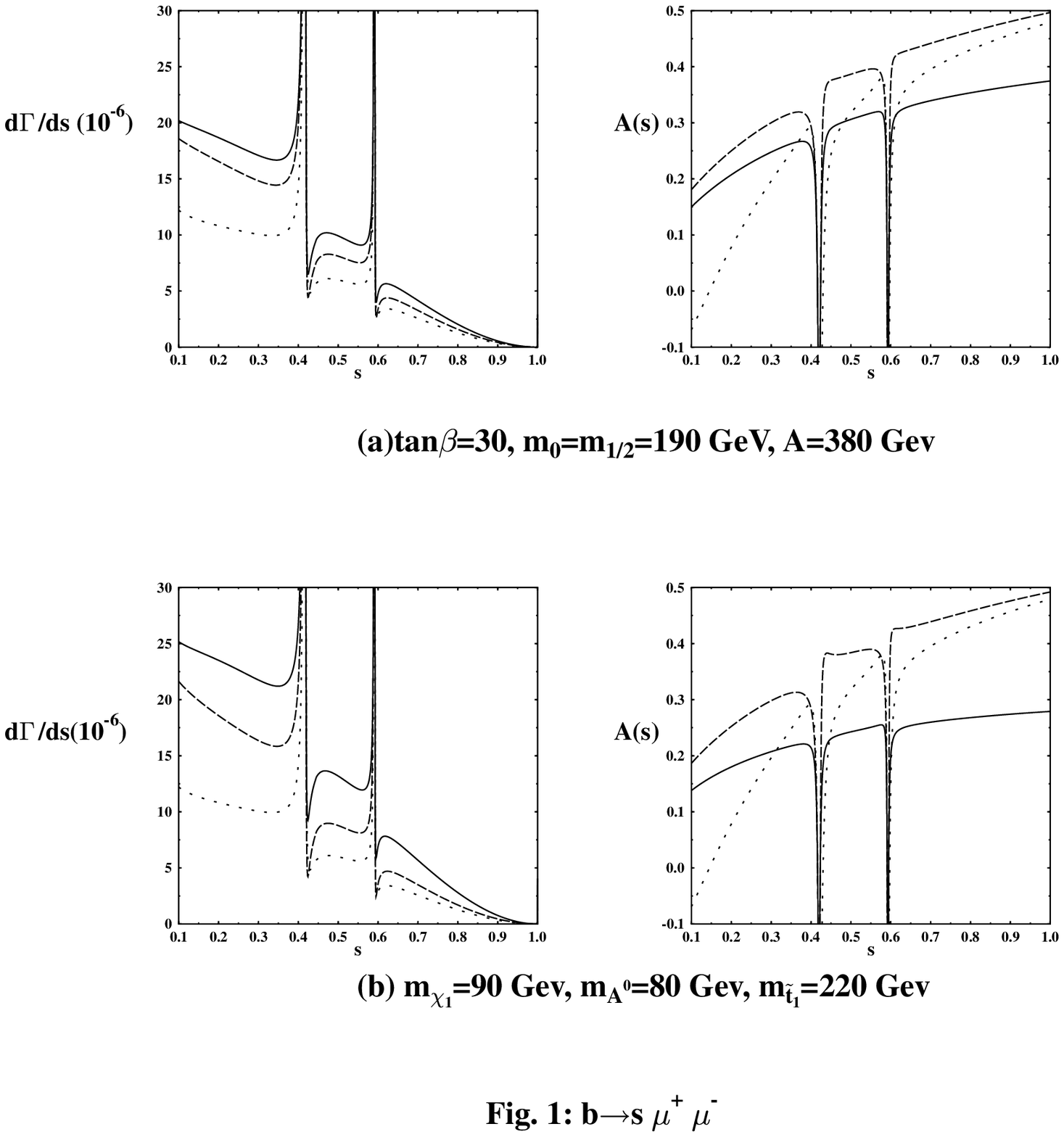,height=8.1 in,rheight=8 in}
  }
\label{fig1}
\end{figure}
\newpage

\begin{figure}[t]
\vspace{0cm}
\centerline{
\psfig{figure=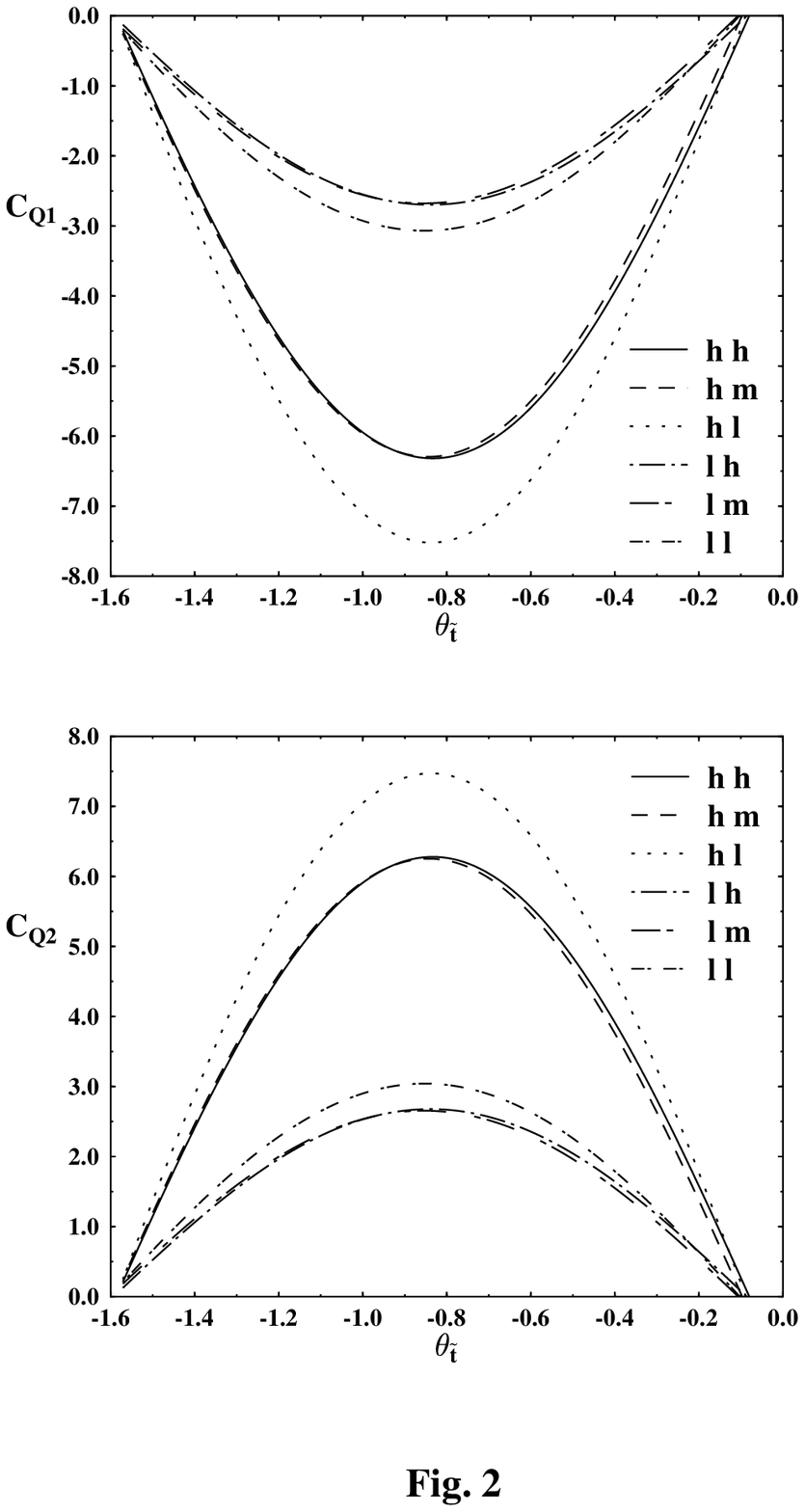,height=8.1 in,rheight=8 in}
}
\label{fig2}
\end{figure}

\end{document}